\documentclass[12pt]{article}
\usepackage{amsmath,amssymb}
\usepackage{graphicx,color,xcolor}
\usepackage{framed,multirow}
\usepackage{paralist,url}
\colorlet{framecolor}{black}
\colorlet{shadecolor}{lightgray}
 \allowdisplaybreaks
\newcommand{\be}{\begin{equation}}
\newcommand{\ee}{\end{equation}}
\newcommand{\bea}{\begin{eqnarray}}
\newcommand{\eea}{\end{eqnarray}}
\newcommand{\bfa}{{\bf a}}

\newcommand{\bfr}{{\bf r}}

\newcommand{\bfv}{{\bf v}}

\newcommand{\bfB}{{\bf B}}

\newcommand{\bfM}{{\bf M}}

\newcommand{\tdD}{{\td D}}

\newcommand{\ds}{\displaystyle}
\newcommand{\nn}{\nonumber}
\newcommand{\pd}{\partial}
\newcommand{\td}{\tilde}
\newcommand{\uA}{{\rm A}}
\newcommand{\uC}{{\rm C}}
\newcommand{\uK}{{\rm K}}
\newcommand{\uN}{{\rm N}}
\newcommand{\uT}{{\rm T}}
\newcommand{\uV}{{\rm V}}
\newcommand{\uW}{{\rm W}}
\newcommand{\um}{{\rm m}}
\newcommand{\us}{{\rm s}}

\newcommand{\nm}{{\rm nm}}
\newcommand{\upm}{{\rm pm}}
\newcommand{\mm}{{\rm mm}}
\newcommand{\cm}{{\rm cm}}
\newcommand{\km}{{\rm km}}
\newcommand{\Hz}{{\rm Hz}}
\newcommand{\mHz}{{\rm mHz}}
\newcommand{\kpc}{{\rm kpc}}
\newcommand{\kg}{{\rm kg}}
\newcommand{\mV}{{\rm mV}}
\newcommand{\pF}{{\rm pF}}
\newcommand{\nW}{{\rm nW}}

\newcommand{\nrad}{{\rm nrad}}
\newcommand{\mK}{{\rm mK}}
\newcommand{\Pa}{{\rm Pa}}
\long\def\symbolfootnote[#1]#2{\begingroup%
\def\thefootnote{\fnsymbol{footnote}}\footnote[#1]{#2}\endgroup}
\begin{document}
%%%%%%%%%%%%%%%%%%%%%%%%%%%%%%%%%%%%%%%%%%%

%%%%%%%%%%%%%%%%%%%%%%%%%%%%%%%%%%%%%%%%%%%
\thispagestyle{empty}
%%%%%%%%%%%%%%%%%%%%%%%%%%%%%%%%%%%%%%%%%%%
%\begin{flushright}
%\hfill{AEI-2014-xxx}
%\end{flushright}
%%%%%%%%%%%%%%%%%%%%%%%%%%%%%%%%%%%%%%%%%%%
\begin{center}

%\vspace{20pt}

{\Large\bf TianQin: a space-borne gravitational wave detector}

\vspace{10pt}

Jun Luo$^1$\symbolfootnote[1]{Email:~\sf junluo@sysu.edu.cn}, Li-Sheng
Chen$^4$, Hui-Zong Duan$^2$, Yun-Gui Gong$^2$, Shoucun  Hu$^6$, Jianghui  Ji$^6$, Qi Liu$^2$, Jianwei Mei$^2$, Vadim Milyukov$^3$, Mikhail Sazhin$^3$, Cheng-Gang Shao$^2$, Viktor T. Toth$^8$, Hai-Bo Tu$^5$, Yamin Wang$^7$, Yan Wang$^2$, Hsien-Chi Yeh$^2$, Ming-Sheng Zhan$^4$, Yonghe Zhang$^6$, Vladimir Zharov$^3$,  Ze-Bing Zhou$^2$

\vspace{10pt}

{\it\small
\baselineskip 20pt
$^1${School of Physics and Astronomy, Sun Yat-Sen University,\\[-5pt]
135 West Xingang Rd., Guangzhou 510275, P.R. China}

$^2${MoE Key Laboratory of Fundamental Quantities Measurement, School of Physics,\\[-5pt]
Huazhong University of Science and Technology, 1037 Luoyu Rd., Wuhan 430074, P.R. China}

$^3${Lomonosov Moscow State University, Sternberg Astronomical Institute, Moscow 119992, Russia}

$^4${Wuhan Institute of Physics and Mathematics, Chinese Academy of Sciences,\\[-5pt]
30 West Xiao Hong Shan, Wuhan 430071, P.R. China }

$^5${Institute of Geodesy and Geophysics, Chinese Academy of Sciences,\\[-5pt]
340 XuDong Rd., Wuhan 430077, P.R. China}

$^6${Key Laboratory of Planetary Sciences, Purple Mountain Observatory, \\[-5pt]
Chinese Academy of Sciences, 2 West Beijing Rd., Nanjing 210008, P.R. China}

$^7${Shanghai Engineering Center for Microsatellites, \\[-5pt]
Building 4, 99 Haike Rd., Shanghai 201203, P.R. China}

$^8${Ottawa, Ontario, K1N 9H5, Canada}
}

\vspace{65pt}
{\bf Abstract}
\end{center}
\vspace{-10pt}

TianQin is a proposal for a space-borne detector of gravitational waves in the millihertz frequencies. The experiment relies on a constellation of three drag-free spacecraft orbiting the Earth. Inter-spacecraft laser interferometry is used to monitor the distances between the test masses. The experiment is designed to be capable of detecting a signal with high confidence from a single source of gravitational waves within a few months of observing time. We describe the preliminary mission concept for TianQin, including the candidate source and experimental designs. We present estimates for the major constituents of the experiment's error budget and discuss the project's overall feasibility. Given the current level of technology readiness, we expect TianQin to be flown in the second half of the next decade.

%%%%%%%%%%%%%%%%%%%%%%%%%%%%%%%
 \newpage
% \setcounter{footnote}{0}
% \setcounter{page}{1}
%%%%%%%%%%%%%%%%%%%%%%%%%%%%%%%

\tableofcontents
%%%%%%%%%%%%%%%%%%%%%%%%%%%%%%%

\section{Introduction}

The existence of gravitational waves (GWs) is a remarkable prediction of Einstein's general theory of relativity. GWs not only offer an important way to test the foundations of general relativity, they are expected to provide a unique and entirely new way to observe and study the Universe. Although the existence of GWs is confirmed indirectly by binary pulsars \cite{Weisberg2010apj030}, to date GWs have not been directly observed. Current efforts aiming at direct detection of gravitational waves include several on-going laser interferometer projects on the ground \cite{Waldman:2011vg,Degallaix2013ol,Somiya2012cqg007} and various proposals for detectors in space (see, e.g., \cite{LISA,ASTROD,DECIGO,OMEGA,nasarif2012,eLISA}). Pulsar timing arrays are being used to search for GW signals from astrophysical sources \cite{EPTA,Hobbs:2008yn,NANOGrav,IPTA}, and the cosmic microwave background is being investigated for signatures of primordial GWs  \cite{Ade:2015tva}. In addition, several upgraded versions of the renowned Weber bar detectors have also continued until very recently \cite{AURIGA,Allegro,MiniGRAIL}.

In this paper, we discuss a new proposal for a space-borne experiment, TianQin\footnote{In Chinese, TianQin means a musical instrument, namely a zither, in space. The TianQin experiment is metaphorically seen as a zither that is being played by the Nature itself through gravitational waves.}, aiming to detect gravitational waves in the millihertz (mHz) range (i.e., $0.1-100$~mHz). GWs in this frequency range could come from a plethora of important astronomical sources, such as ultra-compact galactic binaries, coalescing massive black holes, and from the capture of stellar objects by massive black holes, all of which are exciting. Therefore, in addition to testing gravitational theories under extreme conditions, direct detection of  GWs will also unleash the entirely new field of gravitational wave astronomy.

A unique feature of our proposal is that TianQin's instrument and mission designs, the detector's overall response, as well as the data analysis algorithms will be optimized to detect the GW signal from a single most promising source of gravitational radiation (which we shall call the reference source). Although instrument sensitivity would allow TianQin to detect signals from other sources, accommodating such detections could unnecessarily complicate the instrumental and mission designs,  put demands on the technology and complicate data analysis efforts, thereby increasing the overall mission costs.

Accordingly, instead of designing TianQin as a gravitational wave {\em observatory} capable of studying GWs from a diversity of both known and yet-to-be discovered sources, the design emphasis is placed on the development of a space-borne {\em detector} of gravitational radiation from a single well-understood reference source. The primary goal of TianQin is a direct detection of gravitational waves with anticipated properties from this source \cite{sazhin}, as we will learn how to operate the detector and use it to study the source in great detail.

Our strategy for implementing TianQin is two-pronged. First, we shall identify the strongest source of GWs in the appropriate frequency band and study it with multi-spectral observational tools. Knowledge of the source will help refine the instrumental and mission requirements and focus the overall mission design. Second, we shall, as much as possible, rely on technologies that are readily available or being at advanced stages of development. This will allow us to concentrate our efforts on the development of the two key systems: the laser interferometer and the disturbance reduction system.

For the reference source, there is a set of ultra compact galactic binaries (known as {\em LISA verification binaries} \cite{Stroeer:2006rx}) that have been identified to verify the performance of the proposed eLISA instruments \cite{eLISA}. TianQin is sensitive to the same frequency range, thus the LISA verification binaries are natural candidates as TianQin reference sources. At the moment, RX J0806.3+1527 (also known as HM Cancri or HM Cnc, hereafter J0806) stands out as the best choice, due to its orbital period that is the shortest known to date for a binary system, its relative proximity to the Sun and its moderate component masses.

TianQin is a constellation of three Earth-orbiting spacecraft in a nearly equilateral triangle formation. As shown in later sections, a source like J0806 is strong enough that relatively short arm lengths ($\sim10^5$~km) can be used for the laser interferometer. Compared to heliocentric orbits, the choice of easily accessible geocentric orbits allows the use of more readily available spacecraft technologies, significantly reducing overall mission cost. Most of the existing proposals for space-borne GW antennas have prohibitively high costs, affecting the feasibility of these projects. Our estimates show that the cost for TianQin will be in the more affordable range of USD 550-800 million. This reduced mission cost is made possible by our strategy of building a GW detector rather than an observatory.

Similar to other proposals for space-borne gravitational wave experiments (notably,  \cite{LISA,ASTROD,DECIGO,OMEGA,nasarif2012,eLISA}), TianQin relies on laser interferometry to monitor distance variations between test masses situated inside the spacecraft forming the constellation. The test masses themselves will be subject to various sources of non-gravitational noise, originating both on-board and external to the spacecraft. Thus, a highly efficient disturbance reduction system will be required to reduce the effect of non-gravitational forces on test masses, which will then follow nearly free-fall trajectories, encoding information of transiting gravitational waves. To reduce the contributions of various sources of instrumental noise (primarily laser {frequency} noise), the data analysis will rely on time delay interferometry (TDI, see \cite{Tinto:2014lxa} for a review) implemented on board each spacecraft.

In this paper we present the preliminary concept for TianQin and address the main features of its mission design. This paper is organized as follows: In Section \ref{sec.concept}, we introduce the preliminary concept of TianQin. In Section \ref{sec.goal}, we discuss the sensitivity goal of TianQin and the requirement on the two key components: the laser interferometer and the disturbance reduction system. The projected error budget for these two key components is discussed in Section \ref{sec.budget}, whereas Section \ref{sec.status} offers a brief review of the current status of these key technologies. In Section \ref{sec:summary} we present a discussion of the road map towards implementing TianQin and offer conclusions.

\section{Preliminary mission concept}
\label{sec.concept}

The design for TianQin is guided by the desire to develop an experiment that can be built and launched within a relatively short period of time. An important step towards implementing such an approach is a realistic assessment of the level of the currently available technologies, identification of technology gaps, and initiating efforts to eliminate these gaps. Before discussing the required technologies, we present the current mission concept for TianQin, which relies on three identical spacecraft, placed on nearly identical geocentric orbits with semi-major axis of $\sim10^5$~km, and forming a nearly equilateral triangle. Easily accessible geocentric orbits have been adopted in several other proposals for space-borne gravitational waves observatories. Notably, these include OMEGA \cite{OMEGA,nasarif2012}, LAGRANGE \cite{nasarif2012}, and gLISA \cite{gLISA}. The major advantage of geocentric orbits is a significant reduction in operational costs.

Each spacecraft will be equipped with a laser system capable of sending and receiving laser signals to and from the other two TianQin constellation spacecraft. A heterodyne laser interferometer will be used to monitor distance variations between the spacecraft. Each spacecraft can be chosen as the center spacecraft of the laser interferometer, while the other two become the endpoints of the two interferometer baselines.

\begin{figure}
\begin{center}
\includegraphics[width=13cm,height=6cm]{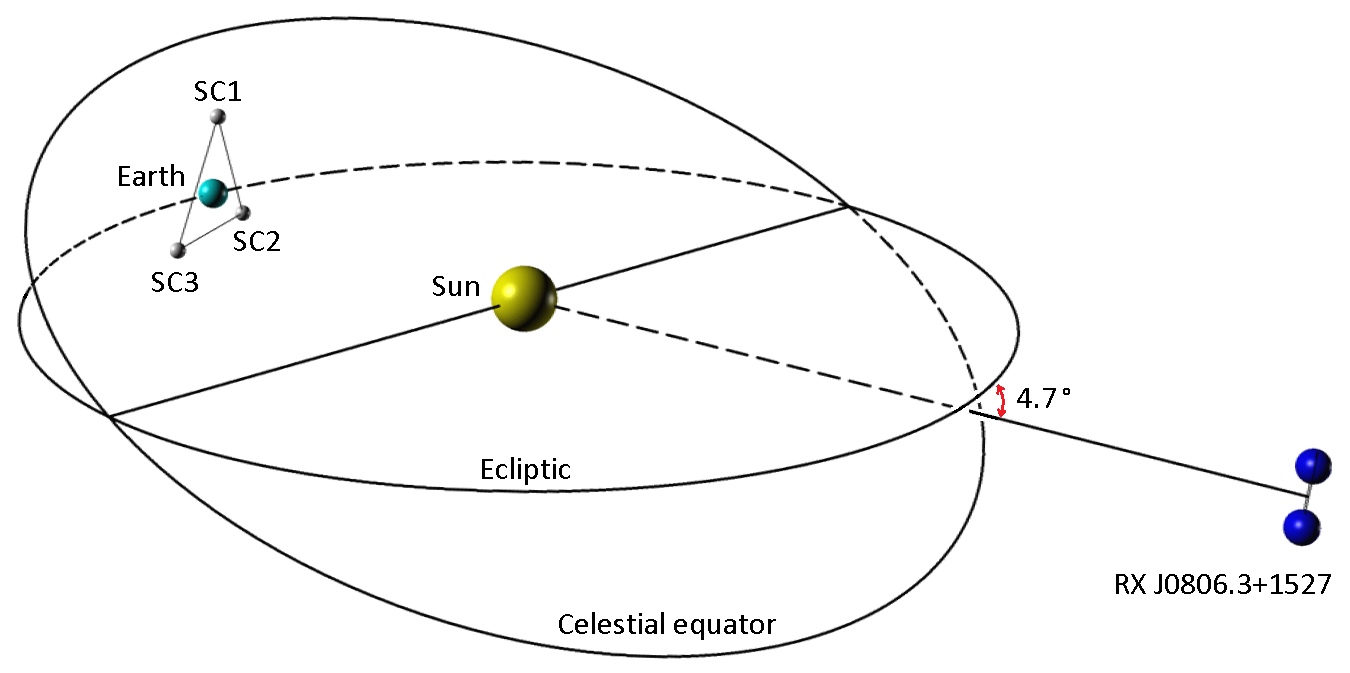}
\caption{An illustration of the preliminary concept of TianQin, with J0806 being the reference source. The three TianQin spacecraft are denoted as SC1, SC2 and SC3. The plane of the celestial equator is also shown, together with the direction to J0806 in the sky.}
\label{fig.tq}
\end{center}
\end{figure}

The spacecraft will each have a disturbance reduction system (DRS) needed to reduce the effects of the  non-gravitational forces on the test masses (which are the reference points for the laser interferometer). As the orbital semi-major axes are $\sim10^5$~km in length, the arm lengths of the laser interferometer are also similar in size. The detector plane is chosen to face the reference source. Narrowband band-pass solar filters will be used to block sunlight from entering the telescopes, and active thermal control is planned to keep thermal fluctuations at the acceptable level. {The orbits of the spacecraft and observational windows will be designed to further reduce the thermal impact on the spacecraft subsystems that may affect the optical path length, e.g., by imposing a sufficient Sun exclusion angle, and if justified, also taking into account effects from Earthshine.} An illustration of TianQin with a tentative reference source is shown in Fig.  \ref{fig.tq}.

TianQin is benefiting from a number of existing proposals for space-borne gravitational waves detectors \cite{LISA,ASTROD,DECIGO,OMEGA,nasarif2012,eLISA}. The extensive literature on LISA verification binaries \cite{Stroeer:2006rx,nelemans} has greatly simplified the task of identifying a possible reference source for TianQin. The use of three identical spacecraft in a nearly equilateral triangle formation, the scheme for the laser interferometer, and many aspects of the DRS closely mimic those of LISA \cite{LISA}. The {proposed} use of {narrowband} band-pass solar filters in OMEGA \cite{nasarif2012} has encouraged us to do the same for TianQin. Apart from these, however, TianQin has its own unique design features.

To simplify the mission design and reduce technological difficulties as much as possible, the sensitivity of TianQin is set to be just enough to detect the most accessible source. The source should be both strong and permanently available, preferably having a short period. (A shorter period can make it easier for the laser interferometer and the inertial sensor to achieve their desired precision goal in the mHz range.) Checked against these requirements, J0806 stands out as the most suitable choice among all the known sources. Therefore, for now we will use J0806 as a tentative reference source to map out the details of various aspects of the experiment. In Sec.~\ref{subsec.j0806} we discuss J0806 in more detail. It should be emphasized, however, that the ultimate choice for the reference source for TianQin is still open and will be refined as time progresses.

We wish to be able to claim a detection within a relatively short period of time, currently identified as approximately three months. This choice of a duration is motivated by an advantageous characteristic of the chosen reference source. As J0806 lies at $\sim 4.7^\circ$ from the plane of the ecliptic, the detector plane is nearly vertical to the ecliptic plane. There are two relatively quiet time windows each year when the direction of the Sun is at a large angle relative to the detector plane (Fig. \ref{fig.tq}). During these times, the amount of sunlight that may enter the telescopes is minimal, which simplifies thermal control. Between these time windows, however, the Sun is near the detector plane, leading to direct sunlight entering the telescopes, causing significant thermal load on the optical system. In this situation, one would have to rely on solar filters and active thermal control to enable continuous science operations of the interferometer throughout the year. Choosing instead a three-month science run, it is possible to operate the constellation only during periods of time when such thermal control is not needed.

The accuracy requirements for the TianQin mission are estimated as $\sqrt{S_x}\sim 1\,\upm/\Hz^{1/2}$ for position and $\sqrt{S_a}\sim 10^{-15}\,\um\,\us^{-2}/\Hz^{1/2}$ for residual acceleration measurements. For a source like J0806 (with parameters given in Table \ref{tab.params}, assuming a distance of $\tdD=5~\kpc$ from the Sun) and arm lengths of $L\sim10^5$~km, a signal-to-noise ratio (SNR) of  10 is possible in three months of integration time.

We must also consider the stability of orbits to be used. For example, the influence of the Moon on the dynamics of orbits becomes more significant as their semi-major axes increase beyond $1\times10^5$~km. We will discuss the chosen orbits in Sec.~\ref{subsec.orbit}.

Although the instruments of TianQin will have the capability to observe sources within a wide range of frequency, at least initially, we will operate the constellation in ``detector mode'', listening for a signal with a known frequency and phase from a pre-selected reference source. When the frequency and phase of the signal that is to be detected are known in advance, a phase-locked detection scheme can be used to help confirm the presence of the signal with high sensitivity. This offers a high degree of confidence that TianQin will achieve its primary mission objective: confirmation of the presence of anticipated mHz gravitational waves by direct detection.

\begin{table}
\begin{center}
\begin{tabular}{c|c}
\hline
Parameter&Value\\
\hline
Number of spacecraft&$N=3$\\
Constellation&Equilateral triangle\\
Type of orbit&Geocentric\\
Arm length&$L\sim 10^5~{\rm km}$\\
Position measurement accuracy&$1~{\rm pm}/{\rm Hz}^{1/2}$~@~6~mHz\\
Residual acceleration accuracy&$10^{-15}~{\rm m}/{\rm s}^2/{\rm Hz}^{1/2}$~@~6~mHz\\
Observation windows&$2\times(3~{\rm months})$ each year\\
Laser wavelength&~~~~~~~~~$\lambda=1064$~nm\\
Optical power&$P_{\rm out}=4$~W\\
Telescope diameter&~~~~~$D=20$~cm\\
Optical efficiency&~$\eta_{\rm opt}=70\%$\\\hline
\end{tabular}
\end{center}
\caption{\label{tbl-params}TianQin basic mission parameters.}
\end{table}

Nonetheless, to achieve some design margin and to leave open the possibility of using TianQin as a GW observatory (perhaps in an extended mission), we aim for ${\rm SNR=10}$ in the construction of the preliminary mission concept.

{A brief summary of the basic TianQin mission parameters is provided in Table~\ref{tbl-params}.}

\subsection{The reference source}\label{subsec.j0806}

J0806 was discovered as a luminous soft X-ray source in the {\it ROSAT} all-sky survey \cite{Beuermann1999trs,Israel2002hcc}. It has been the subject of intense scrutiny in the form of X-ray \cite{Strohmayer2005apj920,Strohmayer2008apjL109} and optical observations \cite{Ramsay2002hc,barros2007mdg,Roelofs2010rms}, due to the fact that, to date, it is the strongest known emitter of periodic GWs in the low frequency band of $10^{-4}-10^{-1}$~Hz, which is accessible to space-based GW detectors.

J0806 presents 100\% X-ray and optical modulations with an apparent period of 321.5~s (5.4 mins). High precision X-ray and optical timing conducted with {\it Chandra} \cite{Strohmayer2005apj920} and {\it Keck-I} \cite{Roelofs2010rms} allowed the determination of the time derivative of the frequency: $\dot{\nu}=3.6\times 10^{-16}{\rm Hz~s}^{-1}$. J0806 is considered a candidate ultra-compact binary white dwarf that can be represented by the ``AM CVn'' model  (semi-detached binary white dwarf) \cite{Solheim2010pasp1133}, with a Roche lobe filling white dwarf losing mass to a more massive white dwarf. In this model, the 5.4 minutes light curve modulation is due to the orbital period of the white dwarf binary.

Other interpretations of this source include the {\it Intermediate Polar} (IP) model and the {\it Unipolar Inductor} (UI) model. The IP model describes the system not as an ultra-compact binary, but rather as a binary system with an orbital period of several hours. The very short signal period actually comes from the spin of magnetic white dwarf. The UI model is essentially a more energetic version of the Jupiter-Io system. A magnetic white dwarf is orbited by a non-magnetic dwarf. In this model, the 5.4 minute period is also the orbital period, but the two stars are detached.

\begin{table}
\begin{center}
{\small
\begin{tabular}{c|c|c|c}
\hline
Parameter  &  Value  &  Uncertainty &  Comments \\\hline
$\nu$ & $3.11\,\mHz$ & $1\times 10^{-10}$ Hz & X-ray/Optical timing  \\
$\dot{\nu}$ & $3.57\times 10^{-16}$ Hz s$^{-1}$ & $2\times 10^{-18}$ Hz s$^{-1}$ & X-ray/Optical timing  \\
$M_{1}$ & 0.55 $M_{\odot}$ & -- & Roche lobe-filling, q, $\dot{\nu}$   \\
$M_2/M_1$ & $0.5$ & $0.13$ & uniform distribution  \\
$\tdD$ & 0.5 to 5\,kpc & -- & X-ray accretion luminosity \cite{Strohmayer2005apj920} and\\
&~&~& temperature of WD components  \\
$\delta$ & $38^{\circ}$ & -- & $M_1$, $M_2$ and $v_{4686}$\\
\hline
\end{tabular}
}
\end{center}
\caption{The system parameters of RX J0806.3+1527. $\nu$ is the orbital frequency, $\dot\nu=d\nu/dt\,$, $M_1$ and $M_2$ are the masses of the two stars, $\tdD$ is distance to J0806, $\delta$ is the inclination, and $v_{4686}$  is the semi-amplitude of the radial velocity for the He II 4686 line. Unless specified otherwise, all data are from \cite{Roelofs2010rms}.}
\label{tab.params}
\end{table}

Using phase-resolved spectroscopy at Keck-I, Roelofs et al. \cite{Roelofs2010rms} have searched for kinematic evidence. They found that the average spectrum of J0806 is dominated by ionized helium emission lines. The full width at half maximum (FWHM) of these lines is about $2500{\rm~km~s}^{-1}$. In the time-resolved spectrum, the He I 4471 line has S-shaped Doppler modulation. Its intensity follows the intensity of the variable continuum flux, which suggests that they originate from the same region. The He II 4686 line is double-peaked and moves in anti-phase with respect to the He I 4471 line. Using the linear back-projection Doppler tomogram, a radial velocity semi-amplitude of $390\pm40{\rm~km~s}^{-1}$ is measured for He I 4471 line, and $260\pm40{\rm~km~s}^{-1}$ for He II 4686 line.

These results favor the AM-CVn model, as illustrated in Fig.~3 of \cite{Roelofs2010rms}. The IP model predicts an hours-long orbital period which is not found in the spectroscopy, and the observed line kinematics do not match the predictions for spectral line variability in the accretor's spin period. The broad and quite constant He II line is a strong signature of accretion, which is not consistent with the UI model.

The system parameters for J0806 are listed in Table~\ref{tab.params}. The parameter with the largest uncertainty is its distance from the Sun. There is approximately a factor of 10 discrepancy between estimates based on X-ray luminosity \cite{Roelofs2010rms} and optical luminosity and temperature \cite{Strohmayer2005apj920}. There is a third, even smaller value (about $0.05$ kpc) listed in \cite{simbad}.

J0806 has a relatively large galactic latitude of about $20^\circ$ \cite{simbad}, suggesting that a distance to J0806 significantly greater than 5~kpc is highly unlikely.

\subsection{Spacecraft orbit}\label{subsec.orbit}

The orbit of a spacecraft is determined by the influence of both gravitational and non-gravitational forces. {The motion of the TianQin spacecraft shall be controlled/reconstructed with an accuracy to the order of $10^{-12}\,\um\, \us^{-2}$ in the relevant frequency band. This will allow the disturbance reduction system (see Sec.~\ref{sec:distred}) to reduce the acceleration noise on the test mass to the order of $10^{-15}\,\um\, \us^{-2}$.}

Important contributions to gravitational forces come from the gravitational field of the extended Earth\footnote{{Higher order multipole moments of the Earth are well known and openly available from projects such as GRACE \cite{GRACE}. It is possible to account for the influence of the multipole moments, including any differential acceleration (which are known to be small) on test masses, in the project design.}} (with relativistic corrections) and the Moon, the monopole gravitational field of the Sun, while the Newtonian monopole gravitational fields of Jupiter, other planets, and the largest asteroids may also contribute. These forces will result in variations of the side-lengths (inter-spacecraft distances) and angles of the triangle formed by the constellation. Distance variations will produce Doppler frequency shifts in the laser interferometer signal and will have to be modeled. Changes in the subtended angles necessitates a pointing control mechanism to align the telescopes of the laser interferometers. The orbits are to be optimized to minimize these variations and their contributions to position and acceleration noise.

\begin{figure}
\begin{center}
\includegraphics[width=16cm,height=9cm]{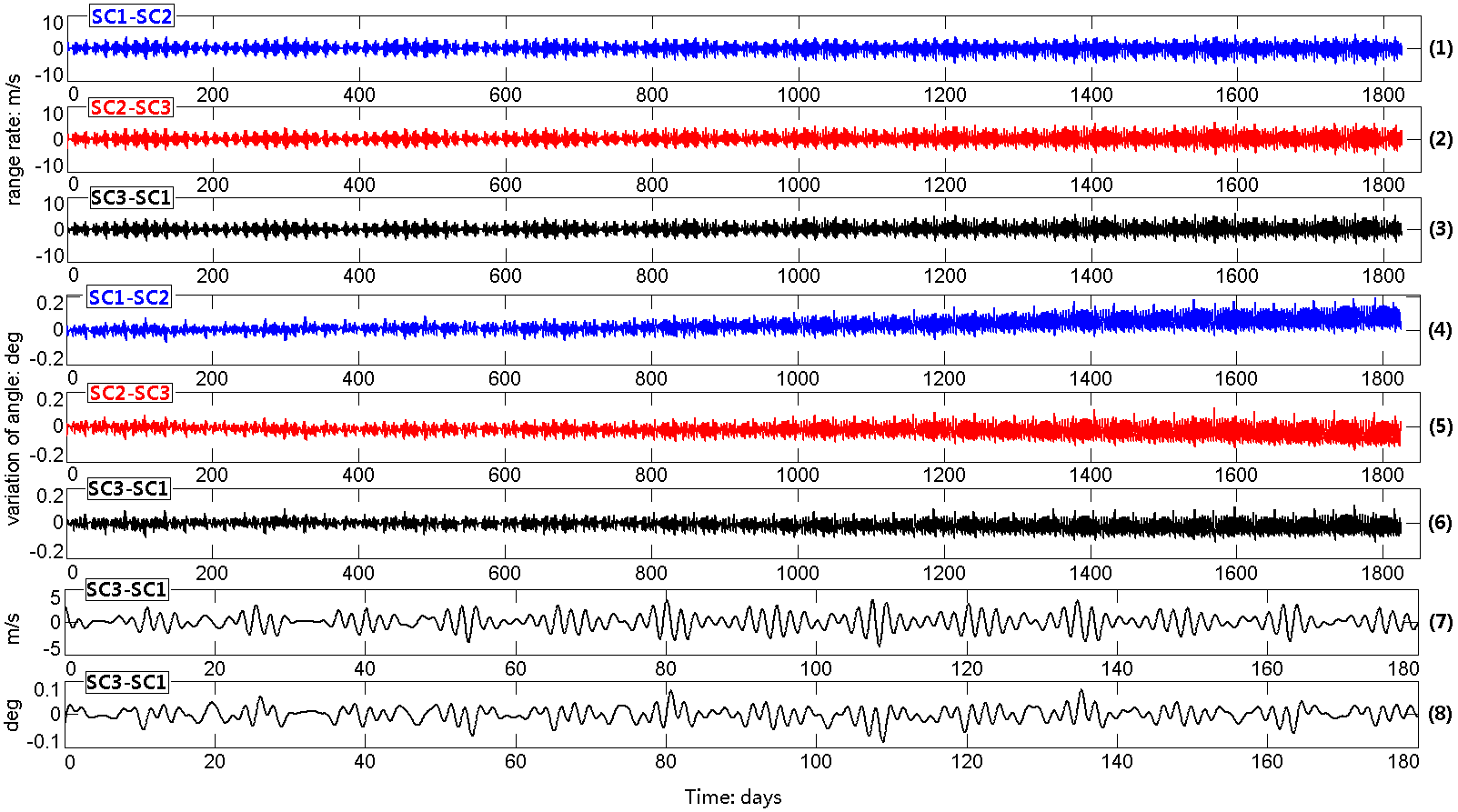}
\caption{Time evolution of the preliminary TianQin orbits. Shown are the range rates (panels 1,2,3 and 7) and the subtended angles (panels 4,5,6 and 8) between each pair of the spacecraft (denoted as SC1, SC2, and SC3) assuming that the three TianQin spacecraft are on nearly identical orbits with a semi-major axis of $10^5$~km. The panels 1--6 show five-year spans, while the panels 7 and 8 show more detail in the first few months for the pair SC3-SC1 (the behavior of the pairs SC1-SC2 and SC2-SC3 is very similar). The effects of the Sun, the Moon and major planets in the solar system, multiple moments of the Earth's gravity up to the fifth order, and random noise at the level $10^{-12} ~{\rm m\, s}^{-2}\,$ (representing the residual effect of non-gravitational forces, which are largely canceled by the drag-free control) have been included in the simulation of the orbits.}
\label{fig.orbitplot}
\end{center}
\end{figure}

A set of possible orbits is shown in Fig.~\ref{fig.orbitplot}. For these orbits, the relative line-of-sight velocity (range-rates) between each pair of spacecraft varies less than 10 m/s over time, and the general behavior remains largely the same for a period of up to five years. The relative velocities will induce Doppler shifts in the laser signals that will have to be compensated or modeled. In the case of LISA \cite{LISA,LISA2000STSR}, the nominal plan is to modulate the laser beams with a signal based on the spacecraft oscillators \cite{Hellings1996oc313,Stebbins1996cqg285}, which allows a range-rate as large as 15 m/s. We will adopt the same approach for TianQin.

The variation of subtended angles can be separated into two parts: short term fluctuations and long-term shifts in the averaged value of the variation. The short term fluctuations have periods on the order of a few days with magnitudes $\sim 0.1^\circ$. This part can be corrected using laser beam pointing control with a fast-steering mirror. On the other hand, attitude control of telescopes using a gimbal mechanism is needed to deal with larger variations that accumulate through a long period of time. This can be done at intervals of the order of several months. {A gimballed telescope has the advantage of lower sensitivity to pointing errors when compared to a fast-steering mirror. The choice between these technologies will be made as the mission requirements are refined and as our understanding of the range of available technological solutions improves.}

To relax the requirement on the disturbance reduction system even further, we may consider implementing a redundant optical truss architecture, as was proposed for the BEACON mission \cite{BEACON}. For this, a fourth spacecraft may be added to the constellation. This option is yet to be investigated.

\section{Sensitivity goal} \label{sec.goal}

The sensitivity of TianQin is set to enable detection of GWs emitted by the chosen reference source. In this study, we are using J0806 as an example that allows us to map out the details of the experiment. Using parameter values from Table~\ref{tab.params}, the GW strain from J0806 is characterized by \cite{MTW1973}
%%%
\be h_0=\frac{2G_N^2M_1M_2}{\tdD\,a}\approx6.4\times10^{-23} \Big(\frac{ M_1}{0.55 M_\odot}\Big)\Big(\frac{M_2}{0.27 M_\odot}\Big) \Big(\frac{ 5\,\kpc}{\tdD}\Big) \Big(\frac{6.6\times 10^4\,\km}{a}\Big)\,,\label{h0}\ee
%%%
with $a$ being the distance between the two stars. (The speed of light is taken to be $c=1$ in this subsection.)

The sensitivity of a Michelson interferometer to gravitational waves is characterized by  (see, e.g., \cite{Larson:1999we,Cornish:2002rt})
%%%
\be h_f=\frac2{\sqrt{R(2\pi f)}}\Big[ \frac{S_x}{L_0^2}+\frac{S_a}{(2\pi f)^4 L_0^2}\Big(1+\frac{10^{-4}\,\Hz}f\Big)\Big]^{1/2} \,,\ee
%%%
where $R(w)$ is the transfer function, $L_0$ is the arm length, while $S_x$ and $S_a$ are the position noise and in residual acceleration noise power densities, respectively. For LISA-like configurations, one has (see, e.g., \cite{eLISA})
%%%
\be R(w)\approx\frac{8}{15}\Big[1+\Big(\frac{w L_0}{0.41\pi}\Big)^2 \Big]^{-1}\,,\label{val.Rw}\ee
%%%
which is obtained by averaging over all sky directions and polarizations, and neglecting small oscillatory behavior in the higher frequency end  \cite{Larson:1999we,Estabrook:2000ef}.

{The transfer function $R(w)$ is most useful for a source with unknown characters and/or located in an unknown sky direction.} For a given binary source {that lies in the optimal direction with respect to the detector plane}, one can define {a specialized} transfer function $R_0(w)$ {that includes} contributions from both polarizations.\footnote{{General transfer functions with the contribution from both polarizations of a gravitational wave have been studied in \cite{Cornish:2002rt}. Here we define $R_0(w)$ following \cite{Larson:1999we}. Further detail can be found in \cite{mei2015ICGAC12}.}} In the low frequency limit, assuming that the detector plane is perpendicular to the incoming gravitational waves, we find
%%%
\be R_0(w)\approx3-2.6\sin^2\delta\,,\label{val.R0}\ee
%%%
which turns out to be independent of the frequency $w\,$. (We will write $R_0(w)$ as $R_0$ from now on.) Apart from the inclination $\delta\,$, other parameters describing the orientation of the binary orbit are often uncertain, and have been averaged over in (\ref{val.R0}).

For J0806, $\delta \approx38^\circ\,$ and $R_0\approx 2.0\,$. The low frequency limit of (\ref{val.Rw}) is $R(w\to0) \approx \frac8{15}\,$, which is about a quarter the optimal value for J0806.\footnote{Since (\ref{val.R0}) is defined for both polarizations put together, while (\ref{val.Rw}) is defined for only one polarization, the total gain in orienting the detector toward J0806 is about a factor 2 improvement over average.}

Assuming that TianQin can conduct effective observations only during the quieter time windows, each observation during such a window yields the integrated strain, from (\ref{h0}),
%%%
\bea h'_0&=&h_0\sqrt{T}\nn\\
&\approx&1.8\times10^{-19}/\Hz^{1/2}\Big(\frac{M_1}{0.55M_\odot}\Big) \Big(\frac{M_2}{ 0.27M_\odot}\Big)\Big(\frac{5\,\kpc}{\tdD}\Big) \Big(\frac{ 6.6\times 10^4\,\km}{a}\Big)\Big(\frac{T}{90{\rm\,days}} \Big)^{1/2}\;.\eea
%%%
Requiring that ${\rm SNR}=10$, we find
%%%
\be\frac{h'_0}{10}\geq \frac2{\sqrt{R_0}}\Big[\frac{S_x}{L_0^2} +\frac{S_a}{(2\pi f)^4L_0^2} \Big(1+\frac{10^{-4}\Hz}f\Big) \Big]^{1/2}\;,\label{sens.requirement}\ee
%%%
which is our basic requirement on the noise goal of the laser interferometer and the disturbance reduction system for TianQin.

For the laser interferometer, we expect to achieve a peak positional sensitivity $\sqrt{S_x}\approx 1$~pm/Hz$^{1/2}$ at $\sim 6\,\mHz$.
The requirement on the residual acceleration $\sqrt{S_a}$ then largely depends on the strength of the source. Assuming $L_0=\sqrt3\,\times10^5$~km
and $\sqrt{S_x}=1$~pm/Hz$^{1/2}$, Eq.~(\ref{sens.requirement}) yields
%%%
\be\frac{S_a^{1/2}}{10^{-15}\,\um\,\us^{-2}/\Hz^{1/2}}\lesssim \left\{~\begin{matrix}34&{\rm when}~\tdD=0.5~\kpc,\cr 17&{\rm when}~\tdD=1~\kpc,\phantom{0.}\cr 3.1&{\rm when}~\tdD=5~\kpc.\phantom{0.}\end{matrix}\right.\ee
%%%
To ensure mission success, TianQin will aim for a successful detection of J0806 in the worst case scenario characterized by $\tdD=5~{\rm kpc}$, by satisfying the most stringent acceleration sensitivity requirement $S_a^{1/2}\approx 10^{-15}\, \um\, \us^{-2}/\Hz^{1/2}$ at $\sim 6\,\mHz\,$.

\begin{figure}
\begin{center}
\includegraphics[width=12cm,height=7.5cm]{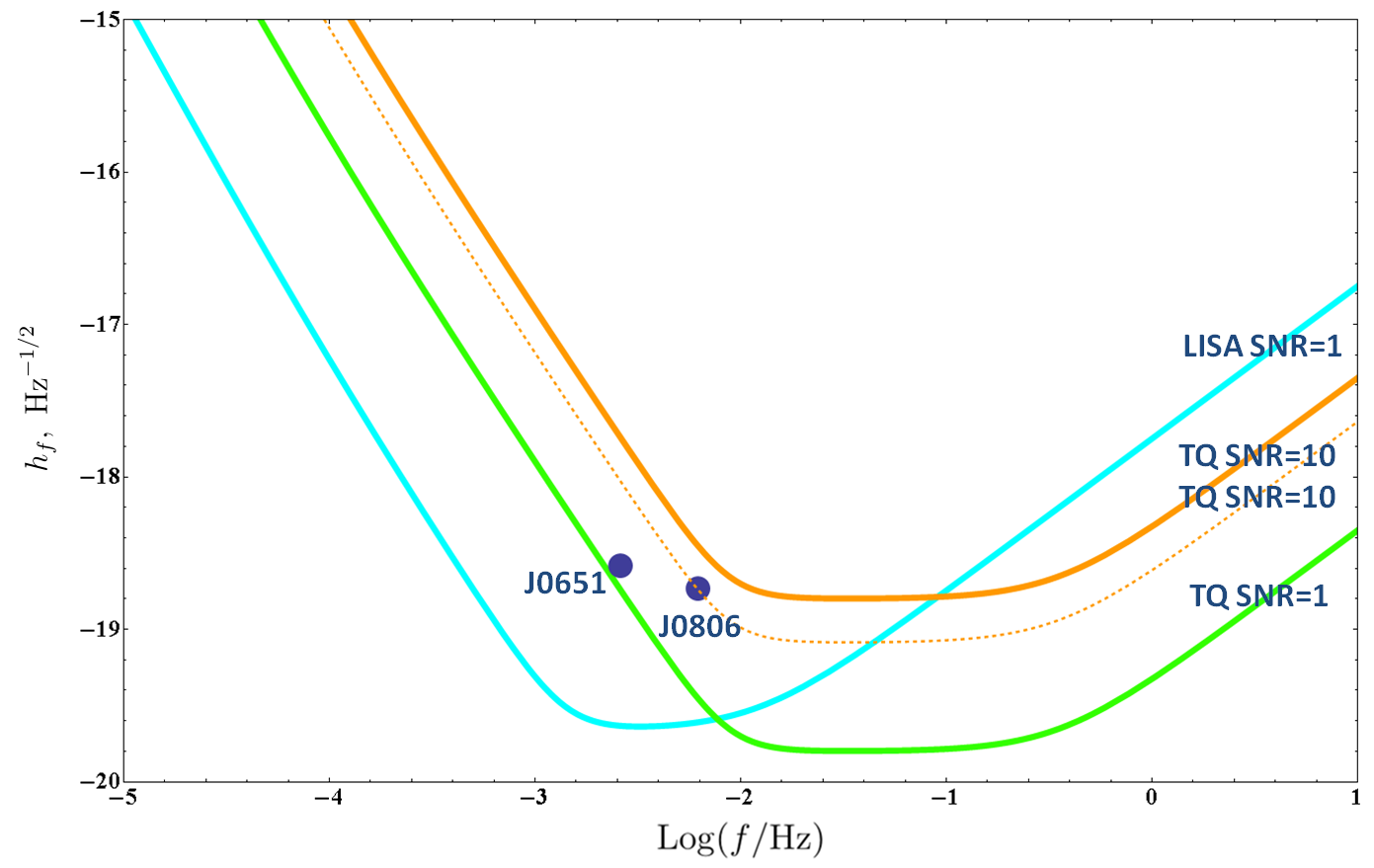}
\caption{The expected sensitivity curve of TianQin. The curve for LISA and another short period binary source, SDSS J065133+2844 \cite{Brown2011apjl23}, are also plotted for comparison. The magnitudes of sources include 90 days of integration time. The solid curves are obtained by using the all-sky and polarization-averaged transfer function (\ref{val.Rw}), while the dashed curve is for sources not only having the same inclination but also lying in the same sky direction as J0806.}
\label{fig.sens-curv}
\end{center}
\end{figure}

The expected sensitivity of TianQin is illustrated in Fig.~\ref{fig.sens-curv}. The gain in optimizing the detector plane towards J0806 is reflected in the distance between the dashed line and the nearby solid curve (both are marked as TQ SNR=10).

\section{Error budget for the key components} \label{sec.budget}

The laser interferometer and the disturbance reduction system are the two key components representing technological challenges. In this section, we discuss a preliminary error budget allocation for each of them.

The preceding section established the position and residual acceleration sensitivity goals of $\sqrt{S_x} =1\,\upm/\Hz^{1/2}$ and $\sqrt{S_a}=10^{-15}\,\um\,\us^{-2}/\Hz^{1/2}\,$, respectively. These requirements are challenging, but there appears to be no fatal obstacle along the path to achieve them.

\begin{figure}
\begin{center}
\includegraphics[width=16cm,height=12cm]{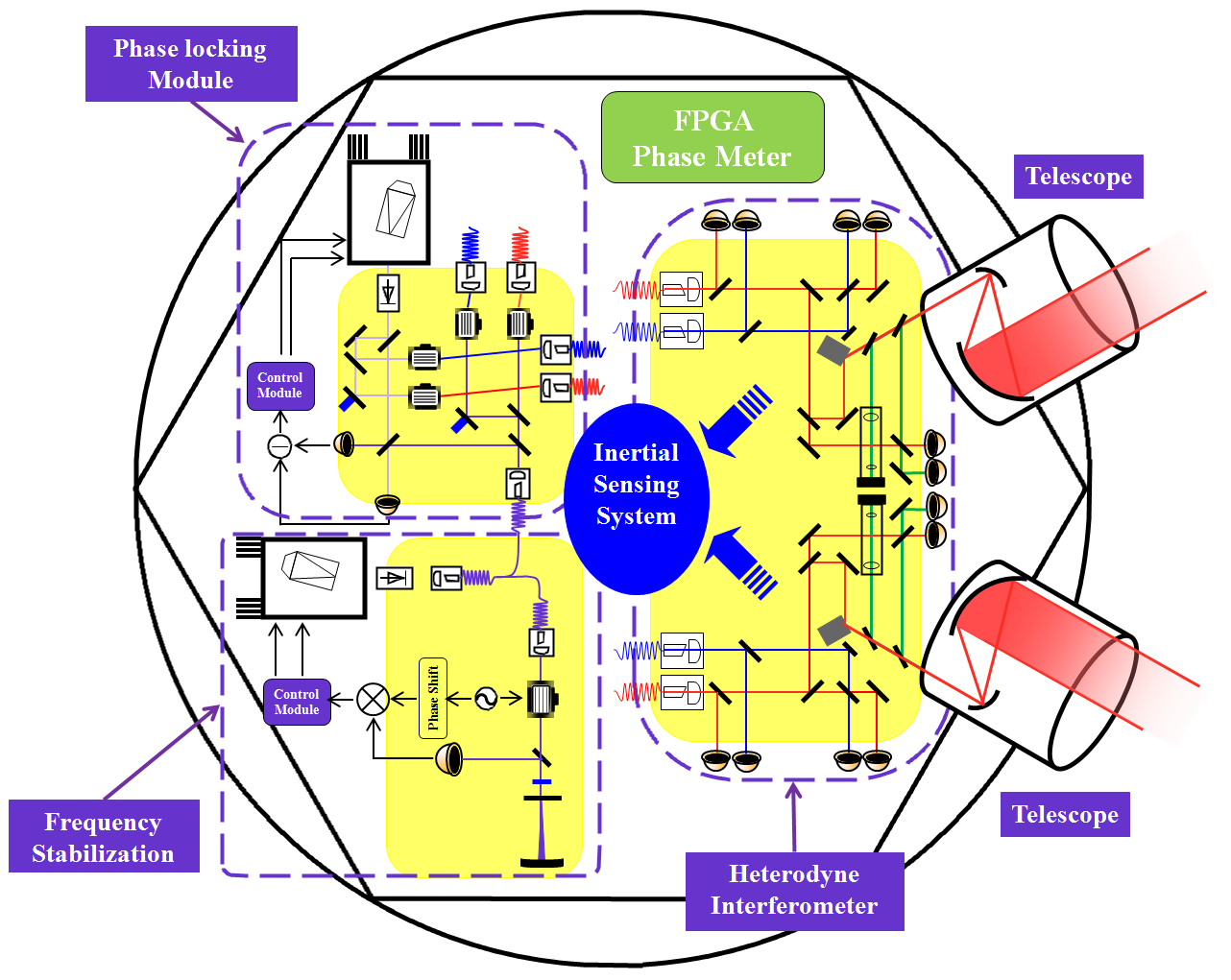}
\caption{A schematic of the optical system on each spacecraft. The system includes a heterodyne interferometer optical bench, a frequency stabilization bench, a phase locking optical bench, an FPGA-based phase-meter and two telescopes. On the interferometer optical bench there are four laser interferometers and a signal acquisition pointing and tracking control optical system. On the frequency stabilization optical bench, there is the Fabry-Perot cavity used to stabilize the laser frequency through the Pound-Drever-Hall scheme. On the phase locking module, the slave laser head is heterodyne optical phase-locked with the frequency stabilized laser head, and two pairs of acousto-optic modulators (AOMs) are used to generate the heterodyne frequencies for heterodyne laser interferometer. All the interfering signals are detected with photo detectors, and we use an FPGA-based ultra high precision phase meter to read the phase of each signal.}
\label{fig.os}
\end{center}
\end{figure}

The scheme for the laser interferometer is very similar to that of LISA \cite{LISA}. On each spacecraft there are three optical modules: modules for laser frequency stabilization and optical phase locking, and the heterodyne interferometer (see Fig.~\ref{fig.os} for details). The laser frequency stabilization module consists of a master laser, an ultra-stable Fabry-Perot cavity and a frequency stabilization control system based on the Pound-Drever-Hall scheme. The optical phase locking module consists of a slave laser and an offset phase lock loop with which the slave laser is phase-locked to the master laser. The heterodyne interferometer module consists of two interferometer optical systems and their corresponding laser beam pointing control systems. Both are bonded on a single piece of ultra-low expansion (ULE) glass baseplate to form one quasi-monolithic optical bench.

Along each of the three arms, one heterodyne transponder-type laser interferometer and two inertial sensing systems will be used to measure the displacement between pairs of caged test masses, one in each spacecraft. The inertial sensing system measures the distance between the test mass and the interferometer optical bench on each spacecraft, while the displacement between the two interferometer optical benches is measured by the transponder-type interferometer.

For the disturbance reduction system, there are two possible schemes for the inertial sensing system. One is to use a spherical test mass with optical readouts in the inertial sensor. In this case, only one inertial sensor will be needed on each spacecraft. The second choice is to use cubic test masses with capacitive sensors in the inertial sensors, just like in LISA \cite{LISA2000STSR}. At present, we leave it open as to which will ultimately be used for TianQin.  Here we will only discuss in detail the scheme with cubic test masses.

In the case with cubic test masses, there are two identical inertial sensors on each spacecraft. Each sensor has a frame of electrodes surrounding a (5~cm)$^3$ cubic test mass, with a capacitive gap of about 5 mm along the sensitive axis. The test masses will be fabricated using an Au-Pt alloy. An audio modulation signal will be injected to the test mass through the injection electrodes on the frame, resulting in a modulated signal. Other electrodes will be used to sense the position of the test mass with respect to the frame and to electrostatic-control the test mass.

In the following subsections, we discuss the main sources of error to the interferometer and the inertial sensors, and then list the preliminary error budget.

\subsection{Laser interferometer}

For the calculations below, we assume the following values: The arm length of the laser interferometer, $L=\sqrt3\times10^5$~km; the laser power, $P_{\rm out}=4\,\uW$; the diameter of telescope, $D=20\,\cm$; the efficiency of the optical chain, $\varepsilon=0.3\,$; the wave-front distortion, $d=\lambda/10\,$; and the central frequency of the laser, $f\approx 2.8\times10^{14}\,\Hz\,$.

The main sources of noise for the heterodyne transponder-type laser interferometer are the following \cite{LISA2000STSR,bender2005cqg339}:

\begin{itemize}

\item {\it Frequency noise:} Laser frequency noise ($\delta f/f$) coupled to the mismatch ($\Delta L$) between the two arm lengths of the interferometer leads to a noise $\delta x_{\rm FN}=(\delta f/f)\Delta L$. Assuming $\Delta L=0.01\,L$, a frequency noise below $\delta f\lesssim 0.1\,\mHz/\Hz^{1/2}$ is required for TianQin. This is to be achieved first by using on-board frequency stabilization system to reach better than $10\,\Hz/\Hz^{1/2}$, and then by using the time-delayed interferometry to reach $0.1\,\mHz/\Hz^{1/2}$.

\item {\it Shot noise:} Quantum fluctuations in the received laser power lead to quantum noise in the phase detection, $\delta x_{\rm QL}=(\lambda/2\pi)\sqrt{h\nu/P_{\rm rec}}$, where $\nu$ is the laser frequency, $\lambda=c/\nu$ is the wavelength and $P_{\rm rec}=\ds\varepsilon P_{\rm out}(\pi w_0D/2\lambda L)^2$, with $w_0$ being the radius of the outgoing beam waist. The axis offset (including a DC pointing error $\theta_{\rm DC}$ and the pointing jitter $\delta\theta$) of the outgoing beam can cause a reduction in the received power. For TianQin, requiring $\delta x_{\rm QL}\lesssim 0.5\,\upm/\Hz^{1/2}$ leads to $P_{\rm rec} \gtrsim 20\, \nW\,$.

\item {\it Pointing stability and wavefront distortion:} Since the measured phase signal is averaged over the detector surface, wavefront distortion and diffraction (due to the finite diameter of telescopes) will influence the signal, leading to a noise $\delta x_{\rm PS/WD}= \frac{1}{64}(2\pi/\lambda)^2d\,D^2\theta_{\rm DC}\delta\theta\,$. For TianQin, requiring $\delta x_{\rm PS/WD}\lesssim 0.5\,\upm/\Hz^{1/2}$ constrains the DC pointing error to $\theta_{\rm DC}\lesssim 10~\nrad$ and the pointing jitter to $\delta\theta\lesssim 10\,\nrad/\Hz^{1/2}$.

\item {\it Thermal stability:} Temperature fluctuations cause the dimensions and the refractive index of optical materials to change, which leads to noise in the optical path length (OPL). To make the whole system stable, ULE will be used to construct the baseplate of the optical system and S-PHM52 \cite{S-PHM52} type optical glass will be used to construct the baseplate of all components. The coefficients of thermal expansion for these materials are of the order $10^{-8}$/K and $10^{-6}$/K\,, respectively. The total temperature-to-OPL coupling coefficient is expected to be $C_{\rm TOPL}\approx 5~\nm/K$. The TianQin phase noise requirement of $\delta x_{\rm OPL}=C_{\rm TOPL}\delta T_{\rm OB}\lesssim 0.5~\upm/{\rm Hz}^{1/2}$ leads to $\delta T_{\rm OB}\lesssim 0.1~\mK/{\rm Hz}^{1/2}$.

\item {\it Clock stability:} The frequency reference for the heterodyne phase-locked loop is provided by an ultra-stable oscillator (USO), the frequency drift of which leads to phase noise in the final result, $\delta x_{\rm USO} =(\delta F/F)(\lambda F/2\pi\nu)\,$, where $\nu$ is the frequency of gravitational waves. For TianQin ($\nu=6\, \mHz\,$, and assuming {$F=20\,$}MHz), requiring $\delta x_{\rm USO}\lesssim 0.5~\upm/{\rm Hz}^{1/2}$ leads to $\delta F/F\lesssim 10^{-15} ~\Hz^{-1/2}\,$, and the corresponding Allan variance $\sigma_{\rm Allan}=\sqrt{2\nu\ln2}\,(\delta F/F)\lesssim 10^{-16}$. Using inter-spacecraft clock transfer to cancel the clock noise in the measurement data, we expect to lower the requirements on clock stability by several orders of magnitude \cite{LISA2000STSR}.
\end{itemize}

The preliminary error budget for the laser interferometer is presented in Table~\ref{tab.error.budget1}.

\begin{table}
\begin{center}
{\small
\begin{tabular}{c|c|c}
\hline
Type  & Requirements & Position error \\
&(at $6\,\mHz$)& ($\upm/\Hz^{1/2}$) \\
\hline
Laser frequency & $\delta f\lesssim 10\,\Hz/\Hz^{1/2}$ (PDH) & 0.5\\
stabilization & $\delta f\lesssim 0.1\,\mHz/\Hz^{1/2}$ (TDI) &\\
\hline
{Shot noise}&{$P_{\rm out}=4$~W,\quad$D=20$~cm}&{0.5}\\
\hline
{Pointing stability}&{$\theta_{\rm DC}\lesssim 10$~nrad}&{0.5}\\
{and wavefront distortion}&{$\delta\theta\lesssim 10$~nrad/Hz${}^{1/2}$}&\\
\hline
Thermal stability of & $C_{\rm TOPL}\lesssim 5\,\nm/\uK$ & 0.5\\
optical bench & $\delta T_{\rm OB}\lesssim  0.1\,\mK/\Hz^{1/2}$ \\
\hline
Onboard USO & $\delta F/F\lesssim10^{-15}/\Hz^{1/2}$ & 0.5\\
& $\sigma_{\rm Allan}\lesssim10^{-16}$ & \\
\hline
\end{tabular}
}
\end{center}
\caption{Preliminary error budget for laser interferometry.}
\label{tab.error.budget1}
\end{table}

\subsection{Disturbance reduction system}
\label{sec:distred}

In the calculations below, the test mass (TM) is assumed to be made of a Pt-Au alloy, in the shape of a cube with each side measuring $5~\cm$ and a mass of $m_{\rm TM}=2.45~\kg$. For temperature, the nominal value of $T=293~\uK$ will be used when needed. Along the sensitive direction, there are two pairs of frame electrodes facing the TM. The gap between TM and a frame electrode is $d_{x0}=5~\mm$, and the effective area is $s_0=8.1~\cm^2$, leading to a capacitance $C_0=\varepsilon_0s_0/d_{x0} \approx1.4~\pF$. In drag-free control, the spacecraft mass is taken to be $M=250~\kg$.

The main sources of noise for the disturbance reduction system are the following \cite{LISA2000STSR,Schumaker2003cqg239}:

\begin{itemize}

\item {\it Thermal fluctuation:} Temperature differences can result in several different types of forces acting on the TMs:
\begin{inparaenum}[(1)]
\item the radiometer effect due to collision with gas molecules, $\delta a_{\rm radio}=(PS/2m_{\rm TM})(\delta(\Delta T)/T)$, where $P$ and $T$ are the local pressure and temperature, respectively, $\Delta T$ is the temperature difference along the sensitive direction, and $\delta(\Delta T)$ is its fluctuation;
\item radiation pressure, $\delta a_{\rm radp}=(P_{\rm th} S/2m_{\rm TM})(\delta(\Delta T)/T)$, where $P_{\rm th}=U/3V$ is the radiation pressure, $U$ is the internal energy of the radiation and $V$ is the volume of the cavity; and
\item outgassing due to the release of absorbed gas molecules from material surfaces, $\delta a_{\rm outgas}=[P (1+2\Theta/T)S/2m_{\rm TM}](\delta(\Delta T)/T)$, where $\Theta$ is the effective activation temperature.
\end{inparaenum}
Among these, the outgassing effect is less well known and could be more dangerous. The effect depends on the poorly understood effective activation temperature $\Theta$ of the sensor surface and the gas conductance of the paths in the sensor head. For TianQin, a preliminary estimation suggests $P\lesssim 10^{-6}$ Pa and $\delta(\Delta T)\lesssim 5~\mu\uK/\Hz^{1/2}\,$, assuming $\Theta=5000~\uK$.

\item {\it Magnetic disturbance:} The acceleration due to static magnetic forces is $\bfa_m=\nabla(\bfM_p\cdot\bfB)/m_{\rm TM}$, where the TM magnetic moment $\bfM_p$ consists of the remanence $\bfM_r$ and the inductive magnetic moment $\chi_mV_{\rm TM}\bfB/\mu_0\,$ with $\chi_m$ being the magnetic susceptibility, $\mu_0$ the vacuum permeability and $\bfB$ consisting of the spacecraft magnetic field $\bfB_{\rm sc}$ and the ambient (space) magnetic field $\bfB_{\rm sp}\,$. In calculating the fluctuations, $\nabla\bfB_{\rm sp}\,$, $\delta(\nabla\bfB_{\rm sc})\,$, $\nabla \bfM_r$ and $\delta\bfM_r$ are often negligible. The spacecraft magnetic field can come from permanent magnets $\bfM_s$ used in attitude control or laser frequency stabilization, $\bfB_{\rm sc}\approx(\mu_0/4\pi) (\bfM_s\cdot\nabla)(\bfr/r^3)$ and $|\nabla\bfB_{\rm sc}|\approx3|\bfB_{\rm sc}|/r\,$, where $\bfr$ is the displacement from the TM. For TianQin, a preliminary estimation suggests $\delta|\bfB_{\rm sc}|\lesssim 2\times10^{-7}\uT/\Hz^{1/2}\,$, assuming $|\bfM_s|\lesssim 1\,\uA\,\um^2$ at $r=0.8\,\um\,$, and $\chi_m\lesssim 10^{-5}\,$.

\item {\it Electrostatic force noise:} A net charge will build up on the TM due to cosmic rays. For TianQin, the limit on the residual charge is estimated to be $q_{\rm max}\lesssim 1.7\times10^{-13}\,\uC\,$, corresponding to a TM potential bias $V_{\rm TM,max}=q_{\rm max}/C_{\rm tot}\lesssim 10\,\mV\,$, assuming the total capacitance between the TM and the surrounding conductors to be $C_{\rm tot}=17\,\pF\,$. The charge control of TM is to be achieved through the UV discharge technique, either continuously or at discrete intervals. The charged TM can interact with magnetic fields to generate a Lorentz force $m_{\rm TM}\bfa_{\rm L}=q\bfv \times \bfB\,$. The fluctuation then depends on $\delta\bfB$ and $\delta q=C_{\rm tot} \delta V\,$, where $\delta V$ is the error in the measurement of the potential. The residual charge also couples to the fluctuation of the patch potential, $\delta V_{\rm FE}$, on the frame electrodes, leading to the acceleration noise $\delta a_{\rm fn,x} =2\ds(\pd_xC)V_{\rm PM}\delta V_{\rm FE}/m_{\rm TM}\,$. A preliminary estimation suggests $\delta V\lesssim 0.1\,\mV$ and $\delta V_{\rm FE}\lesssim  100\,\mu\uV/\Hz^{1/2}\,$.

\item {\it Capacitive sensor:} The capacitive sensing noise is expected to be at the level of $\delta C\lesssim 6.9\times 10^{-7}\,\pF/\Hz^{1/2}$ at $6\,\mHz$ for each of the two sensing channels, corresponding to $x_n=(\sqrt2\,\delta C/2C_0) d_{x0}\lesssim 1.7\,{\rm nm}/\Hz^{1/2}$  in position sensing noise along the sensitive axis. The corresponding acceleration noise is $\delta a_{Cn}=|k_e|\,x_n/m_{\rm TM}\,$, where $k_e$ is the negative electrostatic parasitic stiffness caused by the capacitive sensor itself, which is projected to be $k_e=-2.3\times10^{-7}~\uN/\um$.

\begin{table}
\begin{center}
{\small
\begin{tabular}{c|c|c}
\hline
Type  & Requirements & Acceleration error \\
&(at $6\,\mHz$)& ($10^{-15}\,\um\,\us^{-2}/\Hz^{1/2}$) \\
\hline
Thermal fluctuation& $\delta(\Delta T)\lesssim 5\,\mu\uK/\Hz^{1/2}$ & 0.52\\
& $P\lesssim 10^{-6}\,\Pa$ &\\
& (assuming $\Theta=5000\,\uK\,$) &\\
\hline
Magnetic disturbance& $\delta|\bfB_{\rm sc}|\lesssim 2\times10^{-7}\uT/\Hz^{1/2}$ & 0.24\\
& $\chi_m\lesssim 10^{-5}$ &\\
& $|\bfM_s|\lesssim 1\uA\um^2$ at $r=0.8\,\um$ &\\
\hline
Electrostatic force & $q_{\rm max}\lesssim 1.7\times10^{-13}\,\uC$&0.24\\
& $V_{\rm TM,max}\lesssim 10\,\mV$ \\
& $\delta V_{\rm FE}\lesssim 100\,\mu\uV/\Hz^{1/2}$ \\
\hline
Other parasitic noise & & 0.36\\
\hline
Capacitive sensor & $\delta C\lesssim 6.9\times10^{-7}\,\pF/\Hz^{1/2}$ & 0.16\\
& $x_n\lesssim 1.7\,\nm/\Hz^{1/2}$ & \\
& $k_e\lesssim -2.3\times10^{-7}\,\uN/\um$ & \\
\hline
Crosstalk between axes & $f_{\rm on}\lesssim 10^{-10}\,\uN/\Hz^{1/2}$ & 0.41 \\
& $f_c\lesssim 10^{-5}\,$ & \\
\hline
Micronewton thrusters & $F_{\rm max}\lesssim 100\,\mu\uN$ & 0.35\\
& $F_n\lesssim 0.1\,\mu\uN/\Hz^{1/2}$ & \\
& $H_{\rm open}\lesssim 75$ & \\
\hline
Other coupling noise & & 0.33\\
\hline
\end{tabular}
}
\end{center}
\caption{Preliminary error budget for the disturbance reduction system.}
\label{tab.error.budget2}
\end{table}

\item {\it Crosstalk between axes:} The interaction between the TM and the spacecraft in the direction of orthogonal axes can leak into the sensitive axis due to crosstalk between axes, $a_{\rm fn}=f_{\rm on}f_c/m_{\rm TM}\,$, where $f_{\rm on}$ is the residual non-gravitational force in the orthogonal direction and $f_c$ is the coupling factor characterizing the crosstalk between axes. For this reason, drag-free behavior is also necessary in the direction of non-sensitive axes. For TianQin, a plausible bound is $f_{\rm on}\lesssim 10^{-10}~\uN/\Hz^{1/2}$ and $f_c\lesssim 10^{-5}\,$.

\item {\it Micronewton thrusters:} The maximum force $F_{\rm max}$ from the micronewton thrusters should be able to balance the non-gravitational force on the spacecraft, and the force noise $F_n$ of the thruster determines the noise of the drag-free control. The acceleration noise is $a_{\rm thn} =|k_e|x_{\rm thn}/m_{\rm TM}\,$, where $x_{\rm thn}=F_n/[Mw^2 (1+H_{\rm open})]$ is the displacement noise (between the spacecraft and the reference frame) due to the drag-free control loop, while $H_{\rm open}$ is the open loop gain. A larger open loop gain will help suppress the contribution from the thruster force noise, as long as the control is stable. For TianQin, we expect $F_{\rm max}\lesssim 100\,\mu\uN\,$, $F_n\lesssim 0.1\,\mu\uN/\Hz^{1/2}\,$, $x_{\rm thn}\lesssim 4\,\nm/\Hz^{1/2}$ and $H_{\rm open}\lesssim 75$.

\end{itemize}

In addition to the noise sources enumerated above, there are more sources of noise to be considered. For instance, the thermal distortion of the spacecraft will induce a gravitational disturbance affecting the TM, which calls for a careful design and simulation of the spacecraft structure. Recently, we initiated such activities; results will be reported elsewhere.  In addition, one needs to account for a possible presence of other sources of GWs within the bandwidth expected for the reference source. Not only does such a contribution have a low probability, but given the matched-filter approach that will be used to study the reference source, any such contribution would be small and easily accounted for \cite{phi08}. The relevant analysis is ongoing and will be reported elsewhere.

The preliminary error budget of the disturbance reduction system is presented in Table~\ref{tab.error.budget2}. This set of noise constraints represents our assessment of the technology requirements that are necessary to implement TianQin. Given currently available technologies, these requirements do not present significant challenges.

\section{Technology status} \label{sec.status}

In this section, we provide a brief summary of the present status of two key technologies that will be used in the TianQin project: the laser interferometer and the disturbance reduction system.

\subsection{Laser interferometer}

The first prototype heterodyne laser interferometer with a 10~m arm-length was built at Huazhong University of Science and Technology (HUST) in 2010. Preliminary results showed that a resolution of better than 3 nm could be obtained \cite{yeh2011rsi501}. In order to achieve higher measurement precision, we developed FPGA-based digital phasemeters and an ultra-stable optical bench.

The digital phasemeter worked using a phase-locked loop \cite{liang2012rsi110}. According to preliminary tests and analysis, the measurement error of the digital phasemeter came mainly from the sampling-time jitter of the analog-to-digital converter (ADC). To decompress this noise further, we used the pilot tone correction method to compensate for the influence of the sampling-time jitter of ADC \cite{liang2015rsi106}. The noise level of the phase measurement can be as low as $10^{-6}~$rad/Hz$^{1/2}$ at 0.1~Hz, corresponding to a displacement of 0.2~pm/Hz$^{1/2}$.

We built the first prototype heterodyne interferometer using hydroxide-catalysis bonding \cite{elliffe2005cgqs257}. In order to evaluate the performance of this quasi-monolithic optical bench, we used this interferometer to perform closed-loop positioning control, achieving picometer level precision.

In order to realize a transponding laser ranging system, we built a heterodyne optical phase-locked loop (OPLL) in which the slave laser is phase-locked with an offset frequency to the incoming laser beam. The whole system consisted of a heterodyne interferometer system (to be installed on the master satellite) and an optical phase-locked loop system (to be installed on the slave satellite). The power of the laser beam received by the slave satellite was attenuated to 10nW by using a neutral filter. The preliminary result showed that the OPLL could maintain the lock for more than 20 hours with a residual error of less than 1 nm.

We are developing a space-qualified laser frequency stabilization system using the Pound-Drever-Hall method \cite{luo2015rsi501}. Since the stability of the reference Fabry-Perot cavity is the key constraint, we bonded the Fabry-Perot cavity and the mode-matching components directly onto the same baseplate and the entire optical bench was put into a vacuum chamber with thermal shielding. Preliminary results show that the laser frequency noise at 1~Hz could be controlled within 10~Hz/Hz$^{1/2}$. Further improvements are planned in the form of constructing a Fabry-Perot cavity with higher finesse and compensating for the noise caused by fiber-optic components.

\subsection{Disturbance reduction system}

There has been a continuous effort to develop a low-noise inertial sensor at HUST since 2000. A flight model based on capacitive position transducer and electrostatic feedback control techniques, with a cubic TM surrounded by a series of frame electrodes as described in the last section, has been in orbital test since December of 2013.

The capacitive position transducer is based on the design of a differential transformer bridge \cite{hu2014rsi001}. Experimental studies show that a noise level much below $10^{-6}$~pF/Hz$^{1/2}$ in the bandwidth of interest can be reached when the capacitive-inductive bridge is well tuned, meeting the requirement of the TianQin mission. The intrinsic noise for the developed electrostatic feedback actuator has been verified to be in the order of $\mu\uV/\Hz^{1/2}\,$.

An inertial sensor with a target noise level of $10^{-15}\,\um/ \us^2/ \Hz^{1/2}$ is currently under design and investigation. Several torsion pendulum facilities have been built to investigate the performance of electrostatic accelerometers on ground \cite{tu2010cqg016,zhou2005cqg537}. The noise of the inertial sensor along the rotational degree of freedom and the translational degree of freedom have been verified to be $10^{-13}\, \uN\um/\Hz^{1/2}$ and $10^{-11}\,\uN/ \Hz^{1/2}\,$, respectively. A torsion pendulum test bench with  better common-mode rejection ratio is under development to achieve a better noise performance near the mHz frequencies. The patch potential on a Au-coated material has been studied carefully \cite{yin2014prd001}.

Thrusters with $1~\um\uN$ precision are available from several institutes in China. These institutes are now being engaged to start an intense program to accelerate their research on building the micronewton thrusters needed in gravitational wave missions.

Possible ways to further lower the requirement on the disturbance reduction system are also being investigated.

\section{Summary and Outlook}
\label{sec:summary}

The proposed TianQin mission is a space-based detector of gravitational waves in the mHz frequency band.  To launch this mission in the foreseeable future, we choose the most accessible GW source in the mHz band as the reference source, and optimize all the aspects of the experiment using the properties of the anticipated signal from this source. The primary mission goal is to detect a GW signal with known properties. This makes the experiment more a detector, rather than an observatory.

To reduce mission complexity, we choose geocentric orbits. This makes it possible to focus most of our efforts on the development of the laser interferometer and the disturbance reduction system, which are the two key components showing a gap to what is required by the experiment. A preliminary error budget has been given for them in this paper. All numbers will need to be carefully verified and updated as project development progresses. Nevertheless, we are confident that these two critical systems will reach the required maturity in the near future.

Another element that we need to develop is a comprehensive modeling, simulation, and data analysis software system. We already initiated the development of a simulation system. Preliminary results are encouraging and will be published soon.

We have engaged our colleagues in the astronomical community to initiate a campaign to study J0806 in more detail and also to search for other suitable sources for TianQin.

Two contingent experiments are envisioned to help the development of TianQin. The first one relies on a single Earth-orbiting spacecraft at an altitude of 700~km aiming to test the Equivalence Principle with an accuracy down to $10^{-16}\,$ \cite{gao2011cpl28}. This mission will allow us to bring the technologies of laser interferometry and drag-free control to maturity. The second mission relies on two geocentric spacecraft orbiting the Earth at an altitude of 400~km aiming at mapping the global gravitational field of the Earth using laser tracking between the spacecraft \cite{yeh2011rsi501}. Among other technologies, this mission will allow us to bring the technology of the high precision inertial sensor to maturity. Both of these missions will benefit from the specialized set of modeling, simulation, and data analysis software that we began to develop, with TianQin as the ultimate intended user. These efforts will allow us to develop and refine all the technologies needed for TianQin in the near future.

Concluding, we would like to emphasize the fact that TianQin is an affordable space-based experiment with a significant probability of success. With many technologies either already commercially available or being rapidly advanced, and given the planned geocentric orbits, this mission is well-focused and will be capable of direct detection of gravitational waves from a single reference source or a small set of sources. This makes TianQin a scientifically strong and technologically sound candidate for the space-based experiment towards the end of the next decade. Ultimately, TianQin is expected to serve as a spring board for a fleet of future gravitational wave observatories.

\section*{Acknowledgement}

This work was supported in part by the National Natural Science Foundation of China (Grants No. 11235004, 11273068, 11473073, 11475064), the Russian Foundation for Basic Research (Grant No. 15-52-53070-NNSF), the Natural Science Foundation of Jiangsu Province (Grant No. BK20141509), and the Foundation of Minor Planets of Purple Mountain Observatory. The authors thank all participants at the first and second Workshop of TianQin Science Mission for very helpful discussions.

\end{document}